# Observation of Squeezing in Hidden Optical-Polarization States

Ravi S. Singh[1], Gyaneshwar K. Gupta and Lallan Yadava

Department of Physics, DDU Gorakhpur University, Gorakhpur-273009 (U.P.) INDIA

Email: - [1] yesora27@gmail.com

**Abstract:** - The dynamic feature of monochromatic bi-modal chaotic optical field, enriched with orthogonally-polarized basis-modes propagating collinearly and undergoing Degenerate Parametric Amplification, is investigated to demonstrate Squeezing in Hidden Optical-Polarization states. The Variance (Noise) of Hidden Optical-Polarization Parameters showing squeezing therein is numerically studied by identifying a Squeezing – Function. This squeezing in HOPS is seen to depend critically on interaction time and an Onset-time responsible for the same is demarcated.

## 1 Introduction

The concept of Polarization in Optics is a centuries-old concept enunciated by Christian Huygens while investigating birefringence in Quartz crystal [1]. The optical-polarization ensures the transversal nature of light and is demonstrated by temporal evolution of tip of electric field vector (light vector) at spatial point which, in general, traverses an ellipse and, therefore, light is said to be in the states of elliptical polarization. Varying 'ratio of amplitudes' and 'phase-difference' along two orthogonal basis- modes by a combination of wave plates and/or polarizing beam splitter the elliptically polarized light degenerates into linear and circular polarizations. Polarization in Classical Optics is quantitatively characterized by experimentally measureable Stokes Parameters [2, 3]. Stokes Parameters are basically 'second-order correlations' between complex field-amplitudes along orthogonal basis-modes. Several other techniques such as Jones Matrix, Mueller Matrix and Coherency matrix [4-8] are introduced for quantitative investigation of optical-polarization states. Stokes-Parameters have an edge over these techniques since they are, straightforwardly, applied for characterizing polarization properties of non-classical light in Quantum domain. There the very role of complex field-amplitudes is taken by Bosonic annihilation operators. Field-Quantization of electromagnetic radiation field by Dirac [9] and the advent of lasers in 1960s [10-12] revolutionized and enlarged perennial domain of Classical Optics whose offspring are Non-linear Optics and Quantum Optics. Several investigators [13-15] have contributed in clarifying nonlinear and nonclassical (quantum) properties of light interacting with matter. In 1970's Prakash and Chandra[16-18] and Agarwal [19] derived the density operator of Unpolarized light and defined it by demanding invariance of statistical properties derived by the moments of field amplitudes of all orders. Lehner [20] and others [21-22] re-visited the Unpolarized light offering some new insights. Mehta and Sharma [23] define, rigorously, perfect optical polarization but the treatment doesn't provide a prescription for investigating optical-polarization states. Prakash and Singh [24] worked out an optical-polarization operator in terms of the product of Bosonic inverse-annihilation operator [25] and annihilation operator along two orthogonal basis-modes. This optical-polarization operator helps in testing whether a light in any state is perfectly polarized or not by satisfying the Modified Eigen-value equation. Prakash et al. [26] and Singh [27] generalized the concept of optical-polarization by considering bi-modal monochromatic rectilinearly propagating optical field of which 'ratio of amplitudes' and 'sum of phases' rather than 'ratio of amplitudes' and 'difference of phases', as in usual concept of optical-polarization, along two orthogonal basis-modes are non-random parameters and defined it as 'Hidden Polarized states of light'. This Hidden Optical-Polarization state (HOPS) is weird optical-polarization state since it is not characterized by Stokes-Parameters and, hence, Hidden optical-polarization parameters are introduced for its characterization [28]. Recently, Singh and Gupta [29] proposed the design of Phase-Conjugating Mirror Michelson Interferometer for generating HOPS as well as formal experimental set-up for measuring Hidden Optical-Polarization parameters. The possibility of single-photon sources from semiconductor Quantum dot [30-32] has received surge of activities in manipulating optical-polarization states in Single-Photonic regime. Notably, polarization-state of photon, being easily altered by a combination of rotator (anisotropic Quartz Crystal) and phase retarders (wave plates), leads many landmarked experiments in Quantum Optics testing fundamental postulates of Quantum Mechanics such as Bell-inequalities' violation tests [33, 34], quantum tomography [35], quantum cryptography [36, 37], quantum teleportation, entanglement and precision measurement[38-40].

Squeezing of a quantized optical field (light) [41, 42] pertains to the possibility of reducing fluctuations (noise) in various field-parameters below those of vacuum-level or those possessed by optical field at coherent state. The consecutive decades, 1980s and 1990s witnessed noteworthy interests for generating, detecting and investigating properties of variant sorts of squeezing such as quadrature components squeezing, Number phase squeezing, Higher order squeezing[43, 44], Generalized squeezing[45], Amplitude-squared squeezing[46] in processes like propagation through Kerr media[47, 48], degenerate four-wave mixing[49], interaction of 3-level V-shaped atom with bi-modal coherent state[50], $k^{th}$ harmonic generation [51], optical parametric amplification[52] etc.. Recently, topics like optical estimation of squeezing [53], simultaneous squeezing of multi-modes [54], squeezing with strong photon-number oscillations [55] and temporal evolution of squeezed states [56] have received renewed spurts of activities. Moreover, reduction in noise (squeezing) have find notable applications in Quantum metrology in beating the limit set by Heisenberg Uncertainty principle [57-59]. Heisenberg Uncertainty relation provides the minimal limit of fluctuations (noise) among non-commutable set of canonically conjugate variables. Squeezing in fluctuation of one variable would induce increased fluctuation (anti-squeezing) in other conjugate variable of quantized electromagnetic field.

Stokes-Operators in Quantum domain describe the optical-polarization properties. These parameters are well known to follow SU (2) Lie algebra and, hence, simultaneous measurements of them are precluded. The squeezing (noise-reduction) of one stokes parameters below than that at coherent state of optical field is understood as the signature of 'polarization-squeezed state'. To enhance the sensitivity of polarization-interferometer Garnier et al. [60] generated polarization-squeezed beam using an optical-parametric process. Following the scheme proposed by N. Korolkova et al.[61] W. P. Bowen et al. [62] succeeded in generating polarization-squeezed light beam by interfering the orthogonally-polarized quadrature squeezed beams. Recently, the 'polarization-squeezing' is utilized to characterize the continuous-variable polarization entanglement [63].

The present paper is devoted to study the squeezing in Hidden Optical-Polarization States (HOPS). When bi-modal chaotic light preserving orthogonally-polarized photons and moving collinearly undergoes Degenerate Parametric-Amplification, squeezing in HOPS is observed. It is seen that squeezing in HOPS is perceptible when interaction time approaches a certain Critical value showing an Onset-time for Squeezing. Furthermore, numerical study for dynamic behavior of Squeezing is carried out demonstrating meagre dependence on intensities of the two orthogonal modes by which, in turn, one may infer explicitly that Squeezing in HOPS can be obtained on Intense chaotic Optical pump field too.

The paper is organized in four sections. Section 2 introduces HOPS in Classical as well as Quantum regime. Utilizing Glauber-Coherence functions, a quantum criterion for HOPS is derived. Section 3 deals with Hidden Optical-Polarization Parameters and comparison with Stokes Parameters is drawn. Section 4 investigates the demonstration of squeezing in Degenerate Parametric Amplification.

**2 HOPS in Classical and Quantum Optics**

A monochromatic beam of light (optical field) propagating rectilinearly along z-direction, in Classical optics, is governed by Maxwell's Classical Electromagnetic Theory having Vector Potential (analytic signal),

$$\mathcal{A} = \hat{\mathbf{e}}_x A_{0x} \cos(\psi-\varphi_x) + \hat{\mathbf{e}}_y A_{0y} \cos(\psi-\varphi_y),$$

$$= \text{Re}(\hat{\mathbf{e}}_x A_{0x}\ e^{-i(\psi-\varphi_x)} + \hat{\mathbf{e}}_y A_{0y}\ e^{-i(\psi-\varphi_y)}),$$

$$= [\hat{\mathbf{e}}_x \underline{A}_x + \hat{\mathbf{e}}_y \underline{A}_y]\ e^{-i\psi}, \tag{1}$$

where $\psi = \omega t - kz$, Re stands for real part, $i = (-1)^{-1/2}$, $\mathbf{k}$ (= $k\hat{\mathbf{e}}_z$) is propagation vector, and $\hat{\mathbf{e}}_{x,y,z}$ are respective unit vectors along x-, y-, and z- axes. Obviously, vector potential, $\mathcal{A}$ and, hence, the optical field is completely specified by its real transverse-amplitudes, $A_{0x,0y}$ and phase-parameters, $\varphi_{x,y}$. These four parameters ($A_{0x,0y}$; $\varphi_{x,y}$) have, in general, random spatio-temporal variations. Moreover, optical field, Eq. (1) is typical instance of bi-modal optical field because it needs two random complex-amplitudes $\underline{A}_{x,y} = A_{0x,0y} \exp(i\varphi_{x,y})$ in orthogonal basis-modes ($\hat{\mathbf{e}}_{x,y}\ \bar{k}$) for its complete statistical-characterization

In Quantum Optics the optical field, Eq.(1) is quantized utilizing field-quantization technique due to Dirac [9]. The analytic signal of Vector Potential operator is tacitly expressed in terms of positive-frequency vector potential, $\hat{\mathcal{A}}^{(+)}$ and negative-frequency vector potential, $\hat{\mathcal{A}}^{(-)}$ as

$$\hat{\mathcal{A}} = \hat{\mathcal{A}}^{(+)} + \hat{\mathcal{A}}^{(-)} \qquad (2)$$

where $\hat{\mathcal{A}}^{(+)}$ and $\hat{\mathcal{A}}^{(-)}$ is Hermitian conjugate pair (i.e. $\hat{\mathcal{A}}^{(+)} = \hat{\mathcal{A}}^{(-)\dagger}$). The positive-frequency Vector Potential part $\hat{\mathcal{A}}^{(+)}$ [64, 65] is given by

$$\hat{\mathcal{A}}^{(+)} = (\tfrac{2\pi}{\omega V})^{1/2} [\hat{e}_x \hat{a}_x(t) + \hat{e}_y \hat{a}_y(t)]\, e^{-i\psi} \qquad (3)$$

where $\hat{a}_{x,y}$ are Bosonic-annihilation operators recognized as quantized complex amplitudes of electromagnetic oscillators in $(\hat{e}_{x,y}\, \bar{k})$ basis-modes by which spatio-temporal bi-mode can be excited [66], $\omega$ is angular frequency of the optical field and V is the quantization volume.

Mehta and Sharma [23] has provided strict definition of polarized light in Quantum Optics by transforming rectilinearly propagating bi-modal monochromatic light to a linearly-polarized single-mode on passing through compensator and/or rotator (SU(2)-transformation). Such polarized light may be termed as 'truly' single-mode optical field as the signal is absent in orthogonal mode. The usual (ordinary) polarized light is completely determined either by the pair of non-random 'ratio of amplitudes' and non-random 'difference in phases' in orthogonal bases-modes, $(\hat{e}_{x,y}, \bar{k})$ or by a non-random complex parameter defined as Index of polarization [24]. Prakash and Singh [24] deduced an optical-polarization operator which prescribes the Index of polarization for perfectly polarized optical-field states.

Moreover, in Refs.[26, 27] the concept of Hidden Optical-Polarization States (HOPS) has been introduced in which signal is, in general, present in all modes but only one complex amplitude suffices for its complete statistical description. HOPS, may, therefore, be termed as 'essential' single-mode optical-field state. Recently, Singh and Gupta [29] proposed a formal Phase-conjugating Mirrored Michelson Interferometer for generation of HOPS and experimental set up for measuring Hidden optical-polarization parameters to characterize HOPS. Notably, HOPS has non-random 'sum of phases' and non-random 'ratio of real amplitudes' contrary to 'truly' single-mode polarized optical field where non-random 'difference of phases' and non-random 'ratio of real amplitudes' in orthogonally basis modes ($\hat{e}_{x,y}$, $\bar{k}$) served as characteristic polarization-parameters. Besides adopting basis of description ($\hat{e}_x$, $\hat{e}_y$) one may work in a general basis ($\hat{\varepsilon}$, $\hat{\varepsilon}_\perp$). Here $\hat{\varepsilon}$ is complex unit vector, $\hat{\varepsilon} = \varepsilon_x \hat{e}_x + \varepsilon_y \hat{e}_y$, satisfying normalization condition, $\hat{\varepsilon}^* \cdot \hat{\varepsilon} = |\varepsilon_x|^2 + |\varepsilon_y|^2 = 1$. A unit vector orthogonal to $\hat{\varepsilon}$ is given by complex unit vector $\hat{\varepsilon}_\perp$ satisfying $\hat{\varepsilon}_\perp^* \cdot \hat{\varepsilon}_\perp = |\varepsilon_{\perp x}|^2 + |\varepsilon_{\perp y}|^2 = 1$; $\hat{\varepsilon}_\perp \cdot \hat{\varepsilon}^* = \varepsilon_x^* \cdot \varepsilon_{\perp x}^* + \varepsilon_y^* \cdot \varepsilon_{\perp y}^* = 0$, providing $\frac{\varepsilon_{\perp y}}{\varepsilon_{\perp x}} = \frac{\varepsilon_x^*}{\varepsilon_y^*}$, where dot(.) denotes inner product of cartesian vectors. The analytic signal (vector potential), $\mathcal{A}$ of a single-mode polarized optical field in the mode ($\hat{\varepsilon}_0$, $\bar{k}$) is described by,

$$\mathcal{A} = \underline{A}\, e^{-i\psi} \qquad (4)$$

where $\underline{A} = \hat{\varepsilon}_0 \underline{A}$ is the complex amplitude along $\hat{\varepsilon}_0$. Complex amplitudes of optical-field represented by Eq.(4) in the basis ($\hat{\varepsilon}$, $\hat{\varepsilon}_\perp$) are $\underline{A}_{\hat{\varepsilon}} = (\hat{\varepsilon}^* \cdot \underline{A}) = \underline{A}(\hat{\varepsilon}^* \cdot \hat{\varepsilon}_0)$; $\underline{A}_{\hat{\varepsilon}_\perp} = (\hat{\varepsilon}_\perp^* \cdot \underline{A}) = \underline{A}(\hat{\varepsilon}_\perp^* \cdot \hat{\varepsilon}_0)$. Using Eq.(4) one may derive index of polarization in the basis, ($\hat{\varepsilon}$, $\hat{\varepsilon}_\perp$) as

$$p_{(\varepsilon,\varepsilon_\perp)} = \underline{A}_{\hat{\varepsilon}_\perp}/\underline{A}_{\hat{\varepsilon}} = (\hat{\varepsilon}_\perp^* \cdot \hat{\varepsilon}_0)/(\hat{\varepsilon}^* \cdot \hat{\varepsilon}_0), \qquad (5)$$

for usual polarized light. Decomposing complex amplitudes, $\underline{A}_{\hat{\varepsilon}}$ ($\underline{A}_{\hat{\varepsilon}_\perp}$) in terms of real amplitudes $A_{0\hat{\varepsilon}}$($A_{0\hat{\varepsilon}_\perp}$) and phase parameters $\varphi_{\hat{\varepsilon}}$($\varphi_{\hat{\varepsilon}_\perp}$) as $\underline{A}_{\hat{\varepsilon}}$ ($\underline{A}_{\hat{\varepsilon}_\perp}$) = $A_{0\hat{\varepsilon}}$($A_{0\hat{\varepsilon}_\perp}$) exp ($i\varphi_{\hat{\varepsilon}}$($\varphi_{\hat{\varepsilon}_\perp}$)), Eq. (5) yields, (i) non-random 'ratio of real amplitudes', $A_{0\hat{\varepsilon}_\perp}/A_{0\hat{\varepsilon}}$ and, (ii) non-random 'difference in phases', $\varphi_{\hat{\varepsilon}_\perp} - \varphi_{\hat{\varepsilon}}$ in basis-modes of description ($\hat{\varepsilon}$, $\hat{\varepsilon}_\perp$). Thus, usual (ordinary) polarized light is completely determined either by non-random 'ratio of amplitudes' and non-random 'difference in phases' in the orthogonal modes or by a non-random complex parameter, $p_{(\hat{\varepsilon},\hat{\varepsilon}_\perp)}$ defining index of polarization. Parametrizing complex amplitudes, $\underline{A}_{\hat{\varepsilon}}$ ($\underline{A}_{\hat{\varepsilon}_\perp}$) and $\varphi_{\hat{\varepsilon}}$($\varphi_{\hat{\varepsilon}_\perp}$) by introducing new real parameters $A_0$, $\chi$ and $\Delta$ such that $0 \leq A_0$, $0 \leq \chi \leq \pi$, $0 \leq \bar{\varphi} \leq 2\pi$ and $-\pi < \Delta \leq \pi$ as

$$A_{0\hat{\varepsilon}} = A_0 \cos\tfrac{\chi}{2},\ A_{0\hat{\varepsilon}_\perp} = A_0 \sin\tfrac{\chi}{2},\ \varphi_{\hat{\varepsilon}} = \bar{\varphi} - \Delta/2,\ \varphi_{\hat{\varepsilon}_\perp} = \bar{\varphi} + \Delta/2, \qquad (6)$$

where $A_0$ and $\bar{\varphi}$ are random parameters satisfying, $A_0 = (A_{0\hat{\varepsilon}_\perp}^2 + A_{0\hat{\varepsilon}}^2)^{1/2}$, $\chi = 2\tan^{-1}(\frac{A_{0\hat{\varepsilon}_\perp}}{A_{0\hat{\varepsilon}}})$, $2\bar{\varphi} = \varphi_{\hat{\varepsilon}_\perp} + \varphi_{\hat{\varepsilon}}$ and $\Delta = \varphi_{\hat{\varepsilon}_\perp} - \varphi_{\hat{\varepsilon}}$, Inserting Eq.(6) into Eq.(5), one obtains $p_{(\hat{\varepsilon},\hat{\varepsilon}_\perp)} = \frac{\underline{A}_{\hat{\varepsilon}_\perp}}{\underline{A}_{\hat{\varepsilon}}} = \tan\tfrac{\chi}{2}\, e^{i\Delta}$.

Conditions for 'ratio of real amplitudes' and 'difference in phases' pertaining to HOPS may be casted in terms of a non-random complex parameter,

$$p_{h(\hat{\varepsilon}, \hat{\varepsilon}_\perp)} = \underline{A}_{\hat{\varepsilon}_\perp} / \underline{A}^*_{\hat{\varepsilon}} = \tan \frac{\chi_h}{2} e^{i\Delta_h}, \tag{7}$$

where $\chi_h$ and $\Delta_h$ are non-random angle parameters ($0 \leq \chi_h \leq \pi$ and $-\pi < \Delta_h \leq \pi$), defining Index of Hidden Optical-Polarization. Parameterzing real amplitudes and phase parameters,

$$A_{0\hat{\varepsilon}} = A_0 \cos \chi_h/2, \ A_{0\hat{\varepsilon}_\perp} = A_0 \sin \chi_h/2; \ \varphi_{\hat{\varepsilon}} = \varphi + \Delta_h/2, \ \varphi_{\hat{\varepsilon}_\perp} = -\varphi + \Delta_h/2 \tag{8}$$

where $A_0$ and $\varphi$ are random parameters ($0 \leq A_0$, $0 \leq \varphi \leq 2\pi$) satisfying by $A_0 = (A^2_{0\hat{\varepsilon}_\perp} + A^2_{0\hat{\varepsilon}})^{1/2}$, $\chi_h = 2\tan^{-1}(\frac{A_{0\hat{\varepsilon}_\perp}}{A_{0\hat{\varepsilon}}})$, and $2\varphi = -(\varphi_{\hat{\varepsilon}_\perp} - \varphi_{\hat{\varepsilon}})$, $\Delta_h = \varphi_{\hat{\varepsilon}_\perp} + \varphi_{\hat{\varepsilon}}$. The analytic signal of vector potential of HOPS can, then, be written in general basis-modes, $(\hat{\varepsilon}, \hat{\varepsilon}_\perp)$ as

$$\mathcal{A} = [\hat{\varepsilon} \cos \frac{\chi_h}{2} A_0 e^{i\varphi} e^{i\Delta_h/2} + \hat{\varepsilon}_\perp \sin \chi_h/2 \ A_0 e^{-i\varphi} e^{i\Delta_h/2}] e^{-i\psi}. \tag{9}$$

Obviously, Eq.(9) describes an optical-polarized field in which 'difference of phases', $\varphi$ in orthogonal modes is random parameter but its statistical properties are governed by one random complex amplitude or random real amplitudes, $A_0$ and random phase parameter, $\varphi$.

The Glauber coherence functions [64, 65] which describe correlation properties at any spatio-temporal point in Quantum Optics is,

$$\Gamma^{(m_x m_y n_x n_y)} = \text{Tr}[\rho(0) \hat{\mathcal{A}}_x^{(-)m_x} \hat{\mathcal{A}}_y^{(-)m_y} \hat{\mathcal{A}}_x^{(+)m_x} \hat{\mathcal{A}}_y^{(+)m_y}] \tag{10}$$

where $\rho(0)$ is density operator of optical field. Setting the condition on quantized complex amplitudes,

$$\hat{a}_y(t)\rho(0) = p\hat{a}_x(t)\rho(0), \tag{11}$$

where p is index of polarization, and inserting Eq.(11) into (10) one obtain the Glauber functions.

$$\Gamma^{(m_x, m_y, n_x, n_y)} = p^{*m_y} p^{m_y} \Gamma^{(m_x + m_y, 0, n_x + n_y, 0)} \tag{12}$$

which describes 'truly' single-mode optical-polarization state. Clearly, Glauber coherence function, Eq.(12) is determined by p (index of polarization) and one of quantized complex amplitudes $\hat{a}_x(t)$. Notably, since, Eq. (12) gives the coherence function for polarized light, Eq.(11) may be regarded as quantum analogue of classical criterion $\underline{A}_y = p \underline{A}_x$ for optical- polarized field. Similarly having employed the criterion,

$$\hat{a}_y(t)\rho(0) = p_h e^{-2i\omega t} \rho(0) \hat{a}^\dagger_x(t), \tag{13}$$

where $p_h$ is (Hidden index of polarization), and substituting Eq.(13) into Eq.(10), we get coherence functions for 'essential' single-mode Hidden Optical-Polarization state,

$$\Gamma^{(m_x,m_y,n_x,n_y)} = p_h^{*m_y} p_h^{n_y} \Gamma^{(m_x+m_y,0,n_x+n_y,0)} \tag{14}$$

Obviously, Glauber coherence functions, Eq.(14) are governed by $p_h$ and one of the quantized complex amplitudes $\hat{a}_x(t)$. The Eq.(13) may be regarded as quantum criterion for HOPS, quantum counterpart of $\underline{A}_y = p_h \underline{A}_x^*$.

## 3 Hidden Optical Polarization Parameters

Optical-polarization states in Classical Optics is characterized by Stokes Parameters,

$$s_0 = <|\underline{A}_y|^2 + |\underline{A}_x|^2>; \; s_1 = <|\underline{A}_y|^2 - |\underline{A}_x|^2>; \; s_2 + i\,s_3 = 2<\underline{A}_y^* \underline{A}_x> \tag{15}$$

which take the roles of operators, (stokes operators) for characterizing usual optical-polarization state in Quantum Optics and bear relations,

$$\hat{S}_o = \hat{a}^\dagger_y(t)\hat{a}_y(t) + \hat{a}^\dagger_x(t)\hat{a}_x(t)$$

$$\hat{S}_1 = \hat{a}^\dagger_y(t)\hat{a}_y(t) - \hat{a}^\dagger_x(t)\hat{a}_x(t)$$

$$\hat{S}_2 + i\hat{S}_3 = 2\,\hat{a}^\dagger_y(t)\hat{a}_x(t)$$

whose expectation values in the optical-field state characterized $\rho(0)$,

$$s_0 = <\hat{S}_o> = \mathrm{Tr}\,[\rho(0)\,\{\hat{a}^\dagger_y(t)\hat{a}_y(t) + \hat{a}^\dagger_x(t)\hat{a}_x(t)\,\}],$$

$$s_1 = <\hat{S}_1> = \mathrm{Tr}\,[\rho(0)\{\hat{a}^\dagger_y(t)\hat{a}_y(t) - \hat{a}^\dagger_x(t)\hat{a}_x(t)\}],$$

$$s_2 + i\,s_3 = <\hat{S}_2 + i\hat{S}_3> = 2\mathrm{Tr}\,[\rho(0)\,\hat{a}^\dagger_y(t)\hat{a}_x(t)] \tag{16}$$

where $i = \sqrt{-1}$, $<\;>$ provides ensemble average (expectation value) in Classical Optics (Quantum Optics), Tr is trace of parenthesized quantity and $\underline{A}_{x,y}$ ($\hat{a}_{x,y}(t)$) gives Classical (Quantized) complex amplitudes. Taking non-random vanishing angle parameters ($\chi_h = 0 = \Delta_h$) and the basis of description ($\hat{\boldsymbol{\varepsilon}}, \hat{\boldsymbol{\varepsilon}}_\perp$) as the linear-polarization basis ($\hat{\mathbf{e}}_x, \hat{\mathbf{e}}_y$) in Eq. (12) for evaluation of Stokes Parameters, Eq.(15), noting the fact that random variables $\varphi$ has equal probability between 0 to $2\pi$, one obtains,

$$s_0 = A_0^2 \text{ and } s_1 = s_2 = s_3 = 0 \tag{17}$$

Eq.(17), at first glance, demonstrates that the light is in Unpolarized state which is not, clearly, the fact because light is in HOPS. Several authors [67-70] has critically investigated inadequacy of Stokes-Parameters in characterizing optical-polarization state. The polarization properties of such an optical field, Eq. (12) is described by Hidden Optical –Polarization Parameters, defined in Classical optics,

$$h_0 = <|\underline{A}_y|^2 + |\underline{A}_x|^2>,$$
$$h_1 = <|\underline{A}_y|^2 - |\underline{A}_x|^2>,$$
$$h_2 + i\, h_3 = 2<\underline{A}_y\, \underline{A}_x>, \qquad (18)$$

or, in Quantum Optics,

$$\hat{H}_o = \hat{a}^{\dagger}_y(t)\hat{a}_y(t) + \hat{a}^{\dagger}_x(t)\hat{a}_x(t),$$
$$\hat{H}_1 = \hat{a}^{\dagger}_y(t)\hat{a}_y(t) - \hat{a}^{\dagger}_x(t)\hat{a}_x(t),$$
$$\hat{H}_2 + i\,\hat{H}_3 = 2\, e^{2i\omega t}\, \hat{a}_y(t)\hat{a}_x(t),$$

having expectation in optical-field states described by $\rho(0)$,

$$h_0 = <\hat{H}_o> = Tr[\rho(0)\{\hat{a}^{\dagger}_y(t)\hat{a}_y(t) + \hat{a}^{\dagger}_x(t)\hat{a}_x(t)\}],$$
$$h_1 = <\hat{H}_1> = Tr[\rho(0)\{\hat{a}^{\dagger}_y(t)\hat{a}_y(t) - \hat{a}^{\dagger}_x(t)\hat{a}_x(t)\}],$$
$$h_2 + i\, h_3 = <\hat{H}_2 + i\hat{H}_3> = 2\, e^{2i\omega t}\, Tr[\rho(0)\, \hat{a}_y(t)\, \hat{a}_x(t)], \qquad (19)$$

Following the procedure proposed by N. Korolkova et al.[61] W. P. Bowen et al.[63], experimentally, produced the Optical-Polarized squeezed state allowing interference between quadrature squeezed light and measure the fluctuations in Stokes-Parameters. Similarly, Singh and Gupta[29] formally designed experimental setup for production of Hidden Optical-Polarization State of light and also for measuring the Hidden Optical-Polarization parameters. The Hidden Optical-Polarization operators, Eq.(19) obey commutation relations,

$$[\hat{H}_1, \hat{H}_0] = [\hat{H}_1, \hat{H}_2] = [\hat{H}_1, \hat{H}_3] = 0$$
$$[\hat{H}_0, \hat{H}_2] = 2i\hat{H}_3,\; [\hat{H}_0, \hat{H}_3] = 2i\hat{H}_2$$
$$[\hat{H}_2, \hat{H}_3] = 2i\,(1+\hat{H}_0) \qquad (20)$$

having obvious relationship $\hat{H}_1^2 + \hat{H}_2^2 + \hat{H}_3^2 = \hat{H}_0^2 + 2(1 + \hat{H}_0)$ or $\overline{\hat{H}}^2 - \hat{H}_0^2 = 2(1 + \hat{H}_0)$. Comparing Eq.(20) with the SU (2) Lie group algebraic equations of stokes operators,

$$[\hat{S}_0, \hat{S}_1] = [\hat{S}_0, \hat{S}_2] = [\hat{S}_0, \hat{S}_3] = 0; [\hat{S}_1, \hat{S}_2] = 2i\hat{S}_3, [\hat{S}_2, \hat{S}_3] = 2i\hat{S}_1, [\hat{S}_3, \hat{S}_2] = 2i\hat{S}_1 \quad (21)$$

One may take cognizance that hidden-polarization operators $\hat{H}_1$ commutes with all others while $\hat{H}_0$ not (cf. $\hat{S}_0$). Non-commutability of Hidden optical-polarization parameter precludes their simultaneous measurements. Heisenberg Uncertainty Principle ($\Delta\hat{H}_j^2 \Delta\hat{H}_k^2 \geq \left|\frac{1}{2i}\langle[\hat{H}_j \hat{H}_k]\rangle\right|^2$) can be invoked for hidden optical-polarization operators to give uncertainty products,

$$\Delta\hat{H}_0^2 \Delta\hat{H}_2^2 \geq |\langle\hat{H}_3\rangle|^2,$$

$$\Delta\hat{H}_2^2 \Delta\hat{H}_3^2 \geq |\langle\hat{H}_0\rangle|^2,$$

$$\Delta\hat{H}_3^2 \Delta\hat{H}_0^2 \geq |\langle\hat{H}_2\rangle|^2, \quad (22)$$

where $\langle\Delta\hat{H}_j^2\rangle = \langle\hat{H}_j^2\rangle - \langle\hat{H}_j\rangle^2$ is a shorthand notation for the variance (noise) of the parameter $\hat{H}_j$ (j = 0, 1, 2, 3), whose square root provide the uncertainty in hidden optical-polarization parameters.

**4 Observation of Squeezing**

Light is said to be in Hidden Optical-Polarization squeezed states (HOPSS) if the variance (noise) of one or more of the hidden optical-polarization parameters is smaller than those of vacuum or displaced vacuum (coherent states) having least noise. Allowing monochromatic orthogonally polarized bi-modal chaotic optical field propagating collinearly to be incident on crystal with second-order non-linearity undergoing Degenerate Parametric Amplification, the Hamiltonian of the process, in Heisenberg convention, takes the form,

$$H = \omega [\hat{a}^\dagger_x(t)\hat{a}_x(t) + \hat{a}^\dagger_y(t)\hat{a}_y(t)] + k[\hat{a}^\dagger_x(t)\hat{a}^\dagger_y(t)e^{-2i\omega t} + \hat{a}_x(t)\hat{a}_y(t)e^{2i\omega t}]. \quad (23)$$

Glauber and Mallow [71] obtained the exact solution of equations of motion, $\dot{\hat{a}}_{x,y} = [\hat{a}_{x,y}, \hat{H}]$, where over dot (.) represent time-variation and usage of natural convention $c = \hbar = 1$ is adopted, of quantized complex amplitudes $\hat{a}_{x,y}$ as,

$$\hat{a}_x(t) = e^{-i\omega t}(C\,\hat{a}_x - iS\,\hat{a}^\dagger_y),$$

$$\hat{a}_y(t) = e^{-i\omega t}(C\hat{a}_y - iS\,\hat{a}^\dagger_x), \tag{24}$$

where C and S are hyperbolic time-varying functions defined to be $C \equiv \text{Cosh}2kt$ and $S \equiv \text{Sinh}2kt$.

Squeezing in HOPS is seen by demanding inequalities vis-à-vis with Heisenberg Uncertainty Principle,

$$\langle \Delta \hat{H}_0^2(t) \rangle \text{ or } \langle \Delta \hat{H}_2^2(t) \rangle < |\langle \hat{H}_3 \rangle|,$$

or,

$$\langle \Delta \hat{H}_2^2(t) \rangle \text{ or } \langle \Delta \hat{H}_3^2(t) \rangle < |\langle \hat{H}_0 \rangle|,$$

or,

$$\langle \Delta \hat{H}_3^2(t) \rangle \text{ or } \langle \Delta \hat{H}_0^2(t) \rangle < |\langle \hat{H}_2 \rangle|, \tag{25}$$

The expectation values for the dynamical Hidden Optical-Polarization operators, $\hat{H}_0, \hat{H}_1, \hat{H}_2, \hat{H}_3$, when crystal is pumped with bi-modal chaotic field having density operator [72],

$$\rho(0) = \frac{\bar{n}_x^{\hat{N}_x}}{(1+\bar{n}_x)^{1+\hat{N}_x}} \frac{\bar{n}_y^{\hat{N}_y}}{(1+\bar{n}_y)^{1+\hat{N}_y}} |n_x,n_y\rangle\langle n_x,n_y|, \tag{26}$$

where $\hat{N}_{x,y} = \hat{a}^\dagger_{x,y}\,\hat{a}_{x,y}$ are photonic number-operators and $\bar{n}_{x,y}$ are mean photon numbers along bases $(\hat{e}_{x,y}, \bar{k})$, and their variances (noise) are evaluated to yield after simple algebraic manipulation and insertion of Eq.(26) into Eqs.(19) and Eqs.(22) respectively,

$$\langle \hat{H}_0 \rangle = (\mathcal{N}_y + \mathcal{N}_x)\cosh 4kt + 2\sinh^2 2kt, \tag{27}$$

$$\langle \hat{H}_1 \rangle = \mathcal{N}_y - \mathcal{N}_x, \tag{28}$$

$$\langle \hat{H}_2 \rangle = 0, \tag{29}$$

$$\langle \hat{H}_3 \rangle = (1 + \mathcal{N}_y + \mathcal{N}_x)\sinh 4kt, \tag{30}$$

$$\langle \Delta \hat{H}_0^2(t) \rangle = \sinh^2 4kt + 2\cosh 8kt\,(\mathcal{N}_y\mathcal{N}_x)$$
$$- (1 - 2\cosh 4kt)(\mathcal{N}_y + \mathcal{N}_x) - \cosh^2 4kt\,(\mathcal{N}_y + \mathcal{N}_x)^2, \tag{31}$$

$$\langle \Delta \hat{H}_1^2(t) \rangle = \mathcal{N}_y(1 - \mathcal{N}_y) + \mathcal{N}_x(1 - \mathcal{N}_x), \tag{32}$$

$$\langle \Delta \hat{H}_2^2(t) \rangle = 1 + \mathcal{N}_y + \mathcal{N}_x + 2\, \mathcal{N}_y\, \mathcal{N}_x, \tag{33}$$

$$\langle \Delta \hat{H}_3^2(t) \rangle = \cosh^2 4kt + \cosh 8kt\, (\mathcal{N}_y + \mathcal{N}_x + 2\, \mathcal{N}_y\, \mathcal{N}_x) - \sinh^2 4kt\, (\mathcal{N}_y + \mathcal{N}_x)^2. \tag{34}$$

Here $\mathcal{N}_x \equiv \mathcal{N}_x(n_x) = \dfrac{\bar{n}_x^{n_x}}{(1+\bar{n}_x)^{1+n_x}}$ and $\mathcal{N}_y \equiv \mathcal{N}_y(n_y) = \dfrac{\bar{n}_y^{n_y}}{(1+\bar{n}_y)^{1+n_y}}$ are derived functions depending on photon numbers along two orthogonal basis-modes $(\hat{e}_{x,y}, \bar{k})$.

The squeezing associated with variance, $\Delta \hat{H}_2^2$ is characterized by the inequality, $\langle \Delta \hat{H}_2^2(t) \rangle < |\langle \hat{H}_3 \rangle|$, which gives on putting the values from Eqs.(30) and (33),

$$1 + \mathcal{N}_y + \mathcal{N}_x + 2\, \mathcal{N}_y\, \mathcal{N}_x < |\sinh 4kt|\,|1 + \mathcal{N}_y + \mathcal{N}_x|. \tag{35}$$

From the fact that mean photon numbers, $\bar{n}_x$, $\bar{n}_y$ are non-negative integers, Eq.(35) is satisfied if

$$kt > (\tfrac{1}{4}) \sinh^{-1}\!\left(1 + \frac{2\,\mathcal{N}_x\,\mathcal{N}_y}{1+\mathcal{N}_x+\mathcal{N}_y}\right) \tag{36}$$

One may be prompted by inspecting Eq.(36) to define a Squeezing-Function, Sq $(kt, \mathcal{N}_x, \mathcal{N}_y)$ as

$$\text{Sq}\,(kt, \mathcal{N}_x, \mathcal{N}_y) = 1 + \frac{2\,\mathcal{N}_x\,\mathcal{N}_y}{1+\mathcal{N}_x+\mathcal{N}_y} - \sinh 4kt. \tag{37}$$

Thus, one may conclude from Eq.(37) and Eq.(36) that squeezing in HOPS is perceptible provided Squeezing-Function, Sq $(kt, \mathcal{N}_x, \mathcal{N}_y)$ attains negative values. Numerical Calculations [Figs.(1 a), (1 b), (2 a), (2 b)] regarding Squeezing in HOPS demonstrate critical dependence on interaction time. The interaction time, $kt = 0.22$ second [Figs.(1 a), (1 b))] may be termed as Onset-time as squeezing commences after it.. Moreover, Onset-time and Squeezing is obtained irrespective of intensities in the orthogonal basis-modes revealing, explicitly, that after $kt = 0.22$ second Squeezing ocurrs at very Intense Chaotic Optical pump field too.

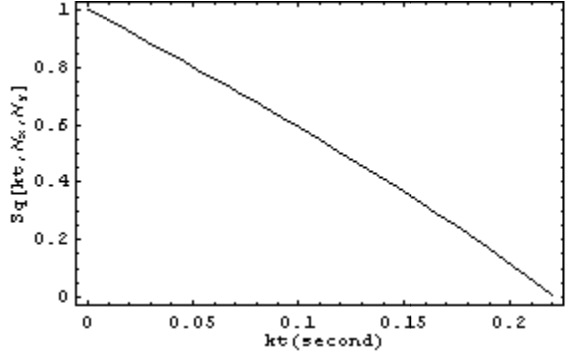

Fig. **(1 a)** Squeezing is not obtained before kt = 0.22 second. The values of $\mathcal{N}_x$ ($n_x$) and $\mathcal{N}_y$ ($n_y$) are 0.035(10) and 0.035(10) respectively having same intensities in bi-modes.

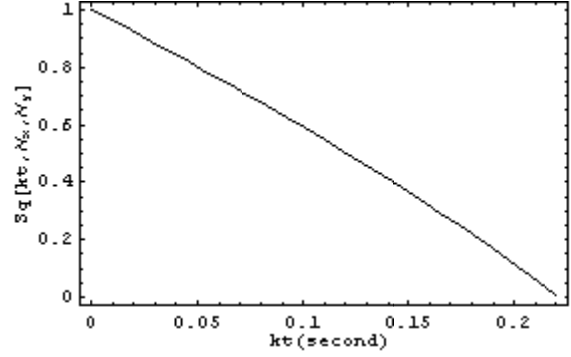

Fig. **(1 b)** Squeezing is not obtained before kt = 0.22 second. The values of $\mathcal{N}_x$ ($n_x$) and $\mathcal{N}_y$ ($n_y$) are 0.25(1) and 0.018(20) respectively having unequal intensities in bi-modes.

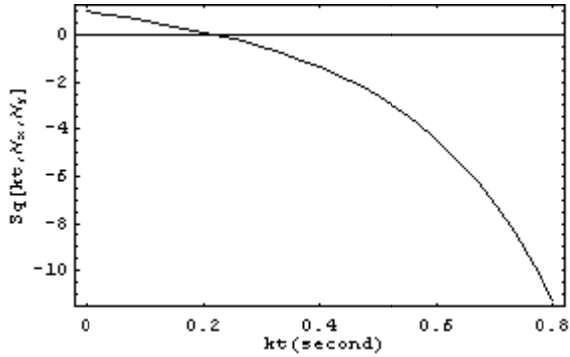

Fig. **(2 a)** Beginning of Squeezing after Onset-time, kt = 0.22 second. The values of $\mathcal{N}_x$ ($n_x$) and $\mathcal{N}_y$ ($n_y$) are 0.035(10) and 0.035(10) respectively having same intensities in bi-modes.

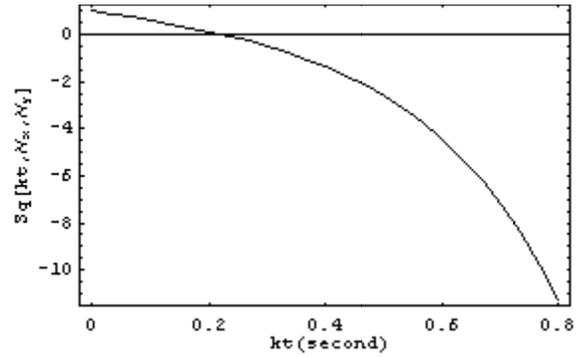

Fig. **(2 b)** Beginning of Squeezing after Onset-time, kt = 0.22 second. The values of $\mathcal{N}_x$ ($n_x$) and $\mathcal{N}_y$ ($n_y$) are 0.25(1) and 0.018(20) respectively having unequal intensities in bi-modes.

**Conclusion: -** Squeezing in Hidden Optical-Polarization is demonstrated by investigating the dynamic behavior of bi-modal chaotic light equipped with orthogonally polarized collinearly propagating photons. A squeezing function is recognized having shown critical dependence on interaction time. An Onset-time kt = 0.22 second is demarcated before which no squeezing is obtained. Furthermore, Onset-time and Squeezing is seen not to depend on intensities in orthogonal bi-modes. This study of variances and Squeezing in HOPS may be applied to characterize continuous-variable Hidden polarization entanglement (to be published elsewhere) which paves the way for utilizing HOPS in Quantum Computation and Communication technology.